\documentstyle[aps,prb,epsf,multicol,rotate,bookmath,array]{revtex}
\newcommand{\lrule}{ \end{multicols} \noindent
  \rule{0.5\textwidth}{0.1mm}\rule{0.1mm}{3pt}\newline }
\newcommand{\rrule}{ \noindent \parbox{\textwidth}{
  \hfill\rule[-3pt]{0.1mm}{3pt}\rule{0.5\textwidth}{0.1mm}}
  \begin{multicols}{2}\noindent }
\def\tamma{\tilde \gamma}
\def\Tamma{\tilde \Gamma}
\def\gammaR{\gamma^{R}}
\def\tammaR{\tamma^{R}}
\def\gammaA{\gamma^{A}}
\def\tammaA{\tamma^{A}}
\def\xxK{x^{K}}
\def\txK{\tilde x^{K}}
\def\GammaR{\Gamma^{R}}
\def\TammaR{\Tamma^{R}}
\def\GammaA{\Gamma^{A}}
\def\TammaA{\Tamma^{A}}
\def\XxK{X^{K}}
\def\TxK{\tilde X^{K}}
\def\CP{{\cal P}}

\def\PR{\hat {\cal P}^{R}}

\def\PRP{\check {\cal P}^{(2)}_+}
\def\PRM{\check {\cal P}^{(2)}_-}
\def\PLP{\check {\cal P}^{(1)}_+}
\def\PLM{\check {\cal P}^{(1)}_-}
\def\SSoo{\hat S_{11}}
\def\SSot{\hat S_{12}}
\def\SSto{\hat S_{21}}
\def\SStt{\hat S_{22}}
\def\DSoo{\hat S_{11}^\dagger}
\def\DSot{\hat S_{12}^\dagger}
\def\DSto{\hat S_{21}^\dagger}
\def\DStt{\hat S_{22}^\dagger}
\def\cd{\!\cdot\!}

\def\ev{{\mbox{\boldmath$e$}}}
\def\fv{{\mbox{\boldmath$f$}}}
\def\gv{{\mbox{\boldmath$g$}}}
\def\jv{{\mbox{\boldmath$j$}}}

\def\a0{\gamma_o}
\def\b0{\tilde \gamma_o}

\def\bsp{{{\mbox{\boldmath${\sigma}$}}}}

\def\bvj{{\mbox{\boldmath$j$}}}
\def\bvx{{\mbox{\boldmath$x$}}}

\def\bvR{{\mbox{\boldmath$R$}}}
\def\bvn{\hat {\mbox{\boldmath$n$}}}
\def\bvp{{\hat{\mbox{\boldmath$p$}}}}
\def\ubvp{\underline{\hat{\mbox{\boldmath$p$}}}}
\def\bvpf{{\hat{\mbox{\boldmath$p$}}_{\!f}}}
\def\sbvR{{\mbox{\boldmath \mbox{$\scriptscriptstyle R$}}}}
\def\bvvf{{\mbox{\boldmath$v$}_{\!f}}}
\def\qcgrad{{i\, \bvvf\!\!\cdot\!\!\grad_{\!\!\sbvR}\, }}

\def\ot{\mbox{$\scriptscriptstyle \otimes$}}

\def\Ds{\Delta}
\def\Dt{{\mbox{\boldmath ${\Delta}$}}}
\def\Ss{{\Sigma}_m}
\def\St{{\mbox{\boldmath ${\Sigma}$}}_m}

\def\bvmu{\hat {\mbox{\boldmath ${\mu}$}}}
\def\bvsigma{{\mbox{\boldmath ${\sigma}$}}}

\begin{document}
\draft
\date{\today}

\title{Josephson currents through spin-active interfaces}

\author{Mikael~Fogelstr\"om\cite{caddress}}
\address{Institut f\"ur Theoretische Festk\"orperphysik, Universit\"at Karlsruhe,
D-76128 Karlsruhe, Germany}

\maketitle
\begin{abstract}
The Josephson coupling of two {\it isotropic s-wave} superconductors through a small,
magnetically active junction
is studied. This is done as a function of junction transparency and of the degree of spin-mixing
occurring in the barrier.
In the tunneling limit, the critical current shows an anomalous $T^{-1}$ temperature 
dependence at low temperatures and for certain magnetic realizations of the junction.
The behavior of the Josephson current is governed by Andreev bound states appearing within
the superconducting gap, $\Delta$, and the position of these states in energy is tunable with
the magnetic properties of the barrier. This study is done using the
equilibrium part of the quasiclassical Zaitsev-Millis-Rainer-Sauls
boundary condition for spin-active interfaces and a general solution 
of the boundary condition is found. This solution is a generalization
of the one recently presented by Eschrig [M.~Eschrig, Phys. Rev B {\bf 61}, 9061 (2000)] 
for spin-conserving interfaces and
allows an effective treatment of the problem of a superconductor in proximity to a 
magnetically active material. 
\end{abstract}

\begin{multicols}{2}
\section{Introduction}
If a superconductor is exposed to magnetically active impurities\cite{shiba,BKS77} or materials\cite{BBP82} the
superconducting state is modified. Josephson coupling two superconductors through 
a magnetically active barrier may lead to what is known as a $"\pi"$-junction\cite{BKS77},
a junction for which the ground state has an internal phase shift of $\pi$ between the
superconductors across the barrier.
If the barrier is extended to an S/F/S-structure, a ferromagnetic (F) layer sandwiched between
two superconductors (S), the critical current will oscillate as the thickness of F is varied\cite{BBP82}.
Additionally, the critical current will also depend on the strength of the exchange field in F.
The principal reason for this strong dependence of junction properties 
is a drastic modification of the local superconducting density of states in the contact region to 
the F-layer
\cite{DA85}. 
Scattering of a magnetically active surface or transmission through a ferromagnetic barrier leads
to a depairing of the Cooper-pairs and the creation of surface or layer Andreev states at energies
within the superconducting gap. 
There have been an extensive experimental effort to explore the
physics above by tunneling through magnetic insulator barriers\cite{STA85,MT94},
by probing the proximity effect in S/F-structures\cite{prox}
and constructing S/F-multilayers\cite{CR99} (see also references therein). The problem at hand is quite formidable
since the strength of the exchange energy is for most ferromagnetic materials, like Ni, Co and Fe, a
sizable part of the Fermi energy $(\sim eV)$ while
superconductivity lives on a much smaller energy scale $(\sim meV)$. To overcome the difference
in energy scales the ferromagnetic layer must be extremely small $(\lesssim nm)$ and only recently
have supercurrents been reported in S/F/S junctions by Veretennikov {\it et al} \cite{V00} using
weak ferromagnetic alloys for the F-layer.

To efficiently model the S/F/S-junction there are two main routes of approach. The first is to
assume an extension of a ferromagnetic metal, now characterized by a length and an exchange field, 
separating the two superconductors.  Within this approach both critical current oscillations\cite{BBP82,RAD91} and
the effect of the exchange field on the Andreev bound states\cite{KF90,RAD99} have been studied. The limitation
of the approach is that it is restricted to small exchange fields, i.e. fields that are comparable to the 
superconducting gap. An alternative approach is to treat the ferromagnetic part as a partially transparent
barrier which transmits the two spin projections differently \cite{BKS77,DA85,mrs88,JB95}. 
Using Bogoliubov-deGennes equations and a WKB-approach for the ferromagnetic barrier\cite{JB95},
Josephson current-phase relations\cite{KTYB97} and quasiparticle tunneling\cite{KTYB99,ZV00} have been
studied for both conventional s-wave and unconventional d-wave superconductors.

In this Paper,
I follow the second path making use of the quasiclassical theory appropriate for
describing low-energy phenomena like superconductivity for which considered energies are small
compared to the Fermi energy $E_F$ [\onlinecite{eilenberger68,SR83,alex85}]. Ferromagnetism will
enter as a boundary problem for
the quasiclassical Green's functions $\hat g(\bvpf,\bvR_s;\varepsilon)$ at a semi-transparent
interface separating two conventional s-wave superconductors. There is
no general restriction of validity of the present work to conventional superconductivity. The physics 
revealed in the simplest system proves to be quite rich without
adding properties related to an unconventional pairing state\cite{KTYB97,KTYB99,ZV00}
and therefore s-wave superconductivity
in proximity to ferromagnetism should be studied in its own right.
In section \ref{BCQC}, a general solution of the quasiclassical boundary condition, as posed
by Millis, Rainer and Sauls \cite{mrs88}, is given for equilibrium Green's functions. In section \ref{LDOS}
the local density of states in proximity to a ferromagnetic insulator is discussed. This is an important
step in understanding the  
Josephson coupling between two s-wave superconductors studied as function
of a simple phenomenological two spin-band scattering $\hat S$-matrix. Section \ref{JCPR}
is devoted to the study of the Josephson coupling and maps out regions where
the junction is in a normal "0"-state and where it switches to the "$\pi$"-state.

\section{Boundary conditions for quasiclassical projectors at spin-active interfaces}
\label{BCQC}
Surfaces and interfaces involve energies of order $E_F$ which in
quasiclassical theory\cite{eilenberger68} are integrated out at the onset. This means that
boundary conditions for the quasiclassical Green's function at
surfaces  and interfaces must 
be posed for the full Green's function satisfying the Gor'kov equation.
Resulting
boundary conditions have then to be
energy integrated into their quasiclassical form \cite{buchholtzrainer79}. 
Physical properties of an interface can then be
accounted for by a suitably chosen scattering S-matrix. 
A boundary condition  
for partially transmitting interfaces was first
derived by Zaitsev \cite{zaitsev84} and, independently, by Kieselmann \cite{kieselmann85}.
The boundary condition was later generalized 
by Millis, Rainer and Sauls (MRS)
to include
spin-active interfaces\cite{mrs88}, i.e. interfaces which transmit and reflect 
quasiparticles differently depending on the spin projection. 

Recently, Eschrig \cite{eschrig99} used a projector method to solve
Zaitsev's boundary condition in general. 
The projectors, 
$\check \CP^{(I)}_{\alpha}$, introduced 
relate to the quasiclassical Green's function as
$
\check g^{(I)}=-i \pi (\check \CP^{(I)}_+ - \check \CP^{(I)}_-).
$
The superscripts $(I)$ label the side of the interface, 
the subscripts $\pm$ are directional indices and finally
the "h\'{a}\v{c}ek" denote the Keldysh-matrix structure of the Green's function.
For a full account on the quasiclassical projectors $\check \CP^{(I)}_{\alpha}$
and their parameterization, I refer
the reader to Eschrig's original paper\cite{eschrig99} and in Appendix \ref{QCT} I give 
a brief review of elements of quasiclassical theory used in this paper.
Written in projectors $\check \CP^{(I)}_{\alpha}$ 
equations (63-66) of MRS reads 
\lrule
\be
\begin{array}{lr}
\PRM\,\ot\, \SStt \PRP \DStt \,\ot\, (\PRM-\check 1)
       &=\PRM\,\ot\, \SSto \PLM \DSto\,\ot\, (\check 1-\PRM)\\*[.1truecm]

\PRP\,\ot\, \DStt \PRM \SStt\,\ot\, (\check 1-\PRP)
       &=\PRP\,\ot\, \DSot \PLP \SSot\,\ot\, (\PRP-\check 1)\\*[.1truecm]

(\PLM-\check 1)\,\ot\,\DSoo \PLP \SSoo \,\ot\, \PLM
       &=(\check 1-\PLM)\,\ot\, \DSto \PRM \SSto\,\ot\, \PLM\\*[.1truecm]

(\check 1-\PLP)\,\ot\, \SSoo \PLM \DSoo\,\ot\, \PLP
       &=(\PLP-\check 1)\,\ot\, \SSot \PRP \DSot\,\ot\, \PLP
\end{array}
\label{sabc}
\ee
\rrule
Here, the non-commutative $\ot$-product is a usual matrix product and a folding of internal energies.
The solution of the system of equations (\ref{sabc}) is facilitated by a convenient 
parameterization of $\check \CP_{\alpha}$ in terms 
of four coherence functions
$\gammaR,\tammaR,\gammaA,\tammaA$ and two distribution functions 
$\xxK,\txK$. These six functions
are 2$\times$2 spin matrices and 
superscripts
(R,A,K) stand for Retarded, Advanced and Keldysh. 
The set of functions above 
obey Riccati-like equations that are easier
to handle than the original quasiclassical matrix equation\cite{eschrig99}.
Especially, they fulfill certain stability criteria when integrated for
along trajectories $\bvx$. The functions $\gammaR,\tammaA$ and $\xxK$
are bounded when integrating the along a trajectory 
$(\bvvf(\bvp)\cdot\bvx > 0)$ and functions $\tammaR, \gammaA$ and $\txK$ 
are bounded integrating in the opposite direction $(\bvvf(\bvp)\cdot\bvx < 0)$ [\onlinecite{ph}].
Restating this in context of the interface problem we can always integrate 
up to the barrier obtaining $\gammaR, \tammaA$, $\xxK$ along 
trajectories with $\bvvf(\bvp)\cdot\bvx > 0$, i.e. along the trajectories
$\bvp_1$ and $\bvp_2$ in Figure \ref{Scattering}. Similarly,
$\tammaR, \gammaA$ and $\txK$ are integrated stably along $\bvvf(\ubvp)\cdot\bvx < 0$, 
i.e. $\ubvp_1$ and $\ubvp_2$ in Figure \ref{Scattering}.
For constructing the Greens's function at the surface that fulfills eqs. (\ref{sabc})
we still need the "scattered"
functions $\GammaR, \TammaA$ and $\XxK$ with $\bvvf(\ubvp)\cdot\bvx < 0$
and $\TammaR, \GammaA$ and $\TxK$ with $\bvvf(\bvp)\cdot\bvx > 0$. These
functions have to be solved for
from eqs. (\ref{sabc}). Using the functions $(\gammaR_i,\tammaR_i,\GammaR_i,\TammaR_i)$,
the Retarded projectors are 
constructed 
as
\centerline{
\begin{minipage}[b]{.49\textwidth}
\begin{figure}
\centerline{\epsfxsize=0.8\textwidth{\epsfbox{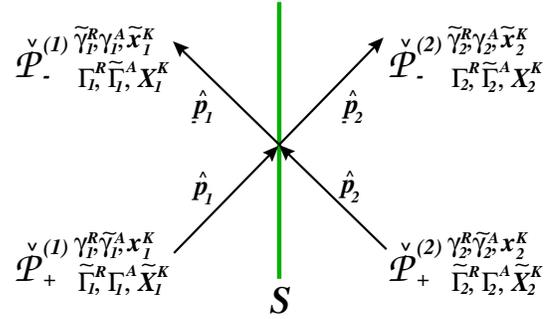}}}
\vspace*{0.5truecm}
\caption[]{
A schematic picture of the in-scattering trajectories $(\bvp_1,\bvp_2)$ 
connected over an interface barrier parameterized by an
S-matrix to the out-scattering ones $(\ubvp_1,\ubvp_2)$}
\label{Scattering}
\end{figure}
\end{minipage}
}
\begin{eqnarray*}
\Bigg \lbrace \begin{array}{l}
\PR_+(\bvp_1) =
\left( \begin{array}{c}1 \\-\TammaR_1\ear \right)
\ot (1 -\gammaR_1\ot \TammaR_1)^{-1} \ot \left(\begin{array}{c} 1 \,\,\, \gammaR_1\ear \right)\\
\PR_-(\ubvp_1) =
\left( \begin{array}{c}-\GammaR_1\\1\ear \right)
\ot (1 -\tammaR_1\ot \GammaR_1)^{-1} \ot \left(\begin{array}{c} \tammaR_1 \,\,\, 1\ear \right)\\
\end{array}\\
\Bigg \lbrace \begin{array}{l}
\PR_+(\bvp_2) =
\left( \begin{array}{c}-\gammaR_2\\1\ear \right)
\ot (1 -\TammaR_2\ot \gammaR_2)^{-1} \ot \left(\begin{array}{c} \TammaR_2 \,\,\, 1\ear \right)\\
\PR_-(\ubvp_2) =
\left( \begin{array}{c}1 \\-\tammaR_2\ear \right)
\ot (1 -\GammaR_2\ot \tammaR_2)^{-1} \ot \left(\begin{array}{c} 1 \,\,\, \GammaR_2\ear \right)\\
\end{array}\\
\end{eqnarray*}
and after substitution into eqs. (\ref{sabc}) one finds after some
straightforward algebra that the scattered-out functions can be expressed solely by 
scattering-in functions as
\be 
\begin{array}{ll}
\GammaR_1&= (S_{11}\gammaR_1 \tilde S_{11}^{-1})\,\,\,\ot\,\, {\cal{R}}^R_{1r}
          + (S_{12}\gammaR_2 \tilde S_{12}^{-1})\,\,\,\ot\,\, {\cal{T}}^R_{1r}\\*[.1truecm]

\TammaR_1&= (\tilde S^\dagger_{11} \tammaR_1 S^{\dagger -1}_{11})  \,\ot\,\, \tilde {\cal{R}}^R_{1r}
          + (\tilde S^\dagger_{21} \tammaR_2 S^{\dagger -1}_{21})  \,\ot\,\, \tilde {\cal{T}}^R_{1r} 
\end{array}
\label{boundarycondition}
\ee
The generalized 
reflection coefficients ${\cal{R}}^R_{Sp}$ are defined
as
\be \begin{array}{l}
{\cal{R}}^R_{1r}=\tilde S_{11} \rho^{R-1}_{21} \,\ot\,
                [\tilde S_{11} \rho^{R-1}_{21}- \tilde S_{12} \rho^{R-1}_{22}]^{-1}\\*[.1truecm]

\tilde {\cal{R}}^R_{1r}=S^\dagger_{11} \tilde \rho^{R-1}_{12} \,\ot\,
                [S_{11}^\dagger \tilde \rho^{R-1}_{12}- S^\dagger_{21} \tilde \rho^{R-1}_{22}]^{-1}
\end{array}
\ee
and corresponding transmission coefficients
${\cal{T}}^R_{Sp}=1-{\cal{R}}^R_{Sp}$. The functions $\rho^R_{ij}$ and $\tilde \rho^R_{ij}$ are
given as
$
\rho^R_{ij}=\tilde S_{ij}-\tammaR_i \ot S_{ij} \gammaR_j\quad\mbox{and}\quad
\tilde \rho^R_{ij}=S^\dagger_{ij}-\gammaR_j \ot \tilde S^\dagger_{ij} \tammaR_i
$. The scattered coherence functions on side 2 are given by interchanging 
the side index $1\leftrightarrow2$.
Advanced functions are related to the retarded ones by general symmetry
$\hat g^A=\hat \tau_3 (\hat g^R)^\dagger \hat\tau_3$.
The similarity in form of the final result given in equation (\ref{boundarycondition})
to the solution of a scattering problem is not a coincidence as the boundary 
condition for the quasiclassical Green's function can also be solved by a direct
scattering approach
\cite{OS00}.

So far no reference to the form of the scattering $S$-matrix has been made. $S$ is
a scalar in Keldysh space and a matrix ${\cal S}$ in particle-hole space, spanned
by Pauli-matrices $\hat \tau_j$, with
the form 
${\cal S} = S (1+\hat \tau_3)/2+\tilde S(1-\hat \tau_3)/2$ where 
$\tilde S(p_{_\parallel})=S^{tr}(-p_{_\parallel})$ [\onlinecite{mrs88}].
For spin-active interfaces the different components of the $S$-matrix, $S_{ij}$
in (\ref{boundarycondition}) above, are $2\times2$ spin matrices. 
To proceed further a specific $S$-matrix is chosen to model the magnetic barrier
\be \hat S=\left(\begin{array}{cr} S_{11} & S_{12} \\ S_{21}& S_{22}\end{array}\right)=
      \left(\begin{array}{cr} r & t \\ t& -r\end{array}\right)\exp( i \Theta \sigma_3)
\label{S-matrix}
\ee
where $\sigma_j$ notes the Pauli-matrices spanning spin space and
parameters $(t,r)$ are the usual transmission and reflection coefficients.
The S-matrix (\ref{S-matrix}) is one of the simplest choices that allows a 
variable degree of spin mixing at the interface
and the spin mixing is parameterized by the spin-mixing angle $\Theta$. 
By this construction $\hat S$ only violates spin conservation, i.e. it does not commute
with the quasiparticle spin operator $\bvsigma$. 
The angle
$\Theta$ will be considered as a phenomenological parameter independent of the trajectory direction
in this paper, but as
shown by Tokuyasu {\it et al} 
in the appendix of Ref. [\onlinecite{TSR88}] one can relate $\Theta$ to the microscopic properties
of the magnetic barrier. In particular, in Ref. [\onlinecite{TSR88}]
an S-matrix is constructed for a magnetically ordered
insulating barrier and it is found, as expected, that $\hat S$ depends
on the quasiparticle momentum projection parallel to the interface and on material parameters 
describing the barrier such as the average band gap, $E_g$, the
internal exchange field, $h_i$, and its orientation $\bvmu$. 

Only the simplest case of an
isotropic s-wave superconductor will be considered in this paper using
weak coupling BCS theory. For this, assuming a constant order parameter in space $\Delta(x)=\Delta$, the 
retarded coherence functions
are $\gammaR=\gamma_0 i \sigma_2$ and $\tammaR=i \sigma_2 \tamma_0$ with 
$\gamma_0=- \Delta/(\varepsilon^R+i \Omega)$, 
$\tamma_0=\Delta^*/(\varepsilon^R+i \Omega)$,
$\Omega=\sqrt{|\Delta|^2-(\varepsilon^R)^2}$ and
$\varepsilon^R=\varepsilon+ i \delta$. If, on the other hand, the effect of proximity to a magnetic material
on the superconductor is of interest the Riccati equations\cite{eschrig99}
\be
\qar
\qcgrad \gammaR+2\varepsilon^R \gammaR = \gammaR \tilde \Delta^R  \gammaR
+\Sigma_d^R \gammaR - \gammaR \tilde \Sigma_d^R-\Delta^R\\
\qcgrad \tammaR -2\varepsilon^R \tammaR = \tammaR \Delta^R \tammaR
+\tilde \Sigma_d^R \tammaR- \tammaR \Sigma_d^R-\tilde \Delta^R
\ear
\label{Riccati}
\ee
have to be solved together with a self consistent determination of
the order parameter $\hat \Delta(x)$ and of the impurity self energy $\hat \Sigma(x;\varepsilon)$ 
as described in the appendix. In eq. (\ref{Riccati}), $\Delta^R=\Delta+\Sigma^R_{od}$ and 
$\tilde\Delta^R=\tilde\Delta+\tilde\Sigma^R_{od}$ are the impurity renormalized gaps, while
$\Sigma_d^R$ and $\tilde\Sigma_d^R$ are diagonal in Nambu space, and include both the 
impurity self energies $\Sigma_{\rm i}$ and the mean fields $\Sigma_{\rm m}$.
All functions entering are spin matrices, and for the problems that will be considered
in this paper it is sufficient to 
parameterize the matrices by two components as
e.g. $(\gamma_o+\gamma_3\sigma_3) i\sigma_2$ and $i\sigma_2(\tamma_0-\tamma_3\sigma_3)$.
The bulk values for $\gammaR$ and $\tammaR$, given above, serve in this case as initial values
when integrating eqs. (\ref{Riccati}).

\section{Andreev bound states at an impenetrable magnetic barrier.} 
\label{LDOS}
As a first point we 
return to the half-space model of Tokuyasu {\it et al}\cite{TSR88}, a semi-infinite
BCS-superconductor bounded by a magnetic insulator. The scattering off the insulator is
assumed to be specular with a phase shift acquired differently at reflection
for spin-up and spin-down
quasiparticles.
Using the $\hat S$-matrix above, the
coherence functions scattered off the magnetic insulator are given
directly by the incoming ones as $\Gamma=\exp{(i \Theta\sigma_3)}
\gamma_0 i \sigma_2$ and $\Tamma =i \sigma_2 \tamma_0\exp{(i \Theta\sigma_3)}$. 
Using the information of the scattered coherence functions the Green's function is given
for a trajectory with $\bvn\cdot\bvpf >0$ as
\be
\hat g^R= -i \pi \; \hat N^R \left(\begin{array}{cc}
(1+\gamma^R \Tamma^R ) & 2\gamma^R
\\
-2\Tamma^R & -(1+\Tamma^R \gamma^R )
\end{array}\right)
,
\label{Pspinpropagator}
\ee
where
\be
\hat N^R= \left(\begin{array}{cc}
(1-\gamma^R \Tamma^R )^{-1} & 0
\\
0& (1-\Tamma^R  \gamma^R )^{-1}
\end{array}\right).
\label{Nfactor}
\ee
If the trajectory is $\bvn\cdot\bvpf <0$, the Green's function is 
simply given by interchanging $\gamma\rightarrow\Gamma$ and
$\Tamma\rightarrow\tamma$.
The effect of the spin
mixing is perhaps best seen in the spin and angle-resolved density of states (DOS) right at the
barrier. On a trajectory with
$\bvn\cdot\bvpf >0$ this quantity  
is given by the imaginary part
of $g^R(\bvpf,\varepsilon)$, the upper left component of (\ref{Pspinpropagator}) 
as described in eq. (\ref{Espindos}).
Assuming
a constant order parameter up to the interface, the local DOS at the interface for spin-up
quasiparticles, $N_{\uparrow}$, can
be written as 
\end{multicols}
\begin{minipage}[t]{0.95\textwidth}
\begin{figure}
\centerline{
\begin{minipage}[t]{0.45\textwidth}
\centerline{\rotate[r]{\epsfxsize=0.7\textwidth{\epsfbox{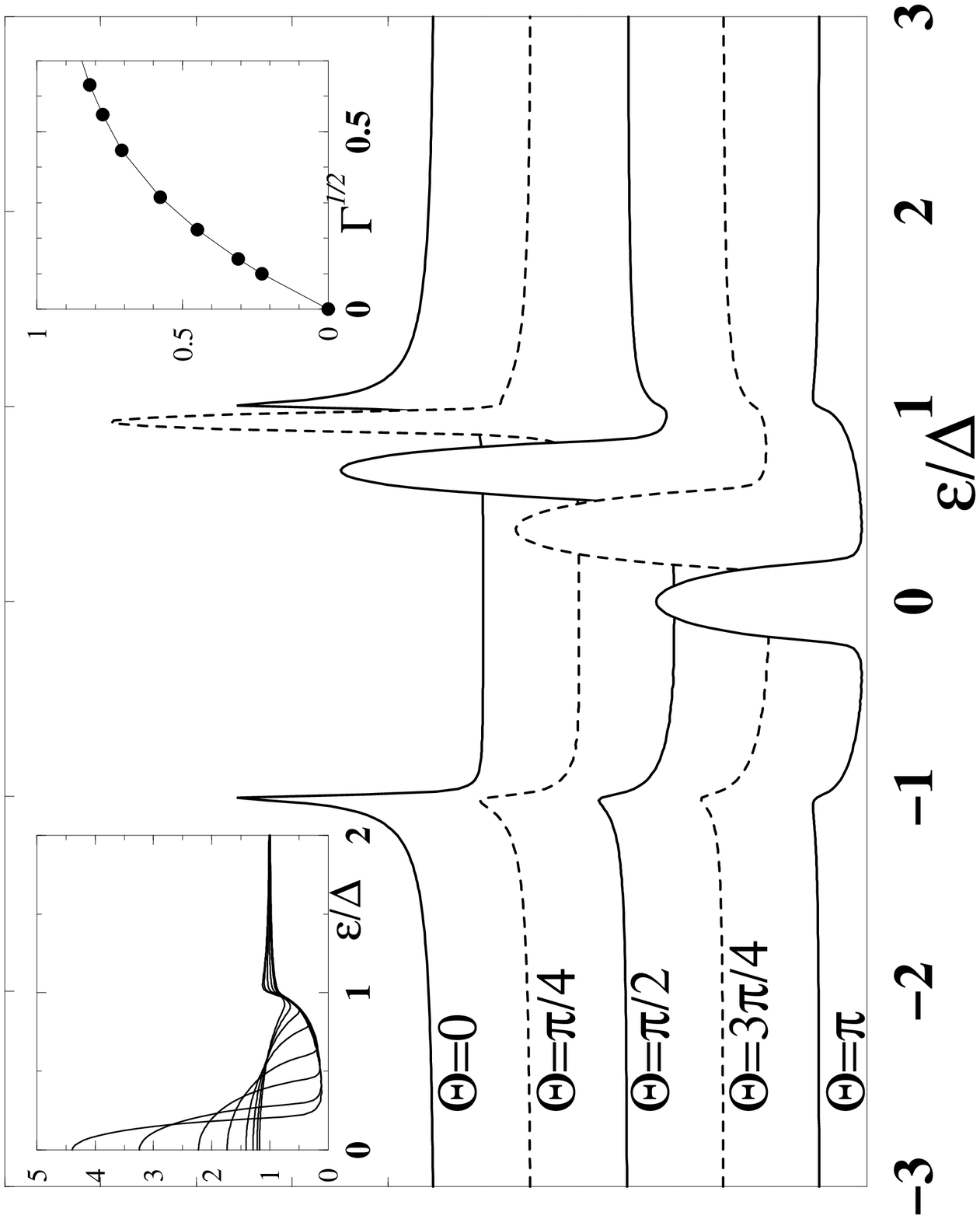}}}}
\end{minipage}
\begin{minipage}[t]{0.45\textwidth}
\centerline{\rotate[r]{\epsfxsize=0.7\textwidth{\epsfbox{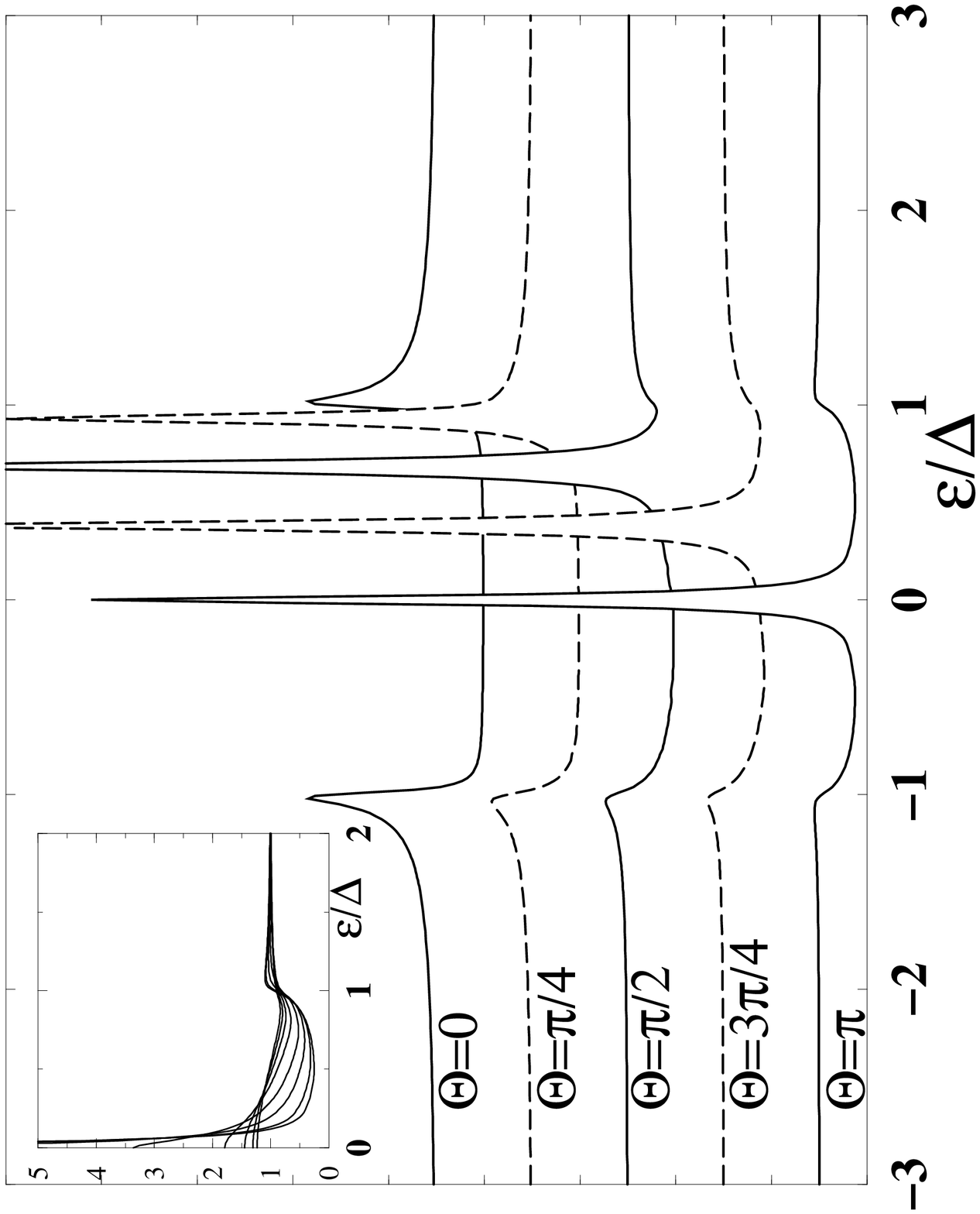}}}}
\end{minipage}
}
\vspace*{0.5truecm}
\caption[]{
Tunneling density of states for spin-up quasiparticles at an impenetrable magnetically active
insulator. The superconductor is a conventional BCS superconductor with a self
consistently determined order parameter. In the main panels $\Theta$ is
varied from 0 to $\pi$ in steps of $\pi/4$ going top to bottom and the curves are shifted for
clarity. In the left figure the impurity scattering is in the Born limit while in the right panel 
the scattering is taken in the unitary limit. In both cases the scattering rate is chosen to $0.01$ in
units of $2\pi T_c$.
In the left inset in the left figure the DOS is plotted for $\Theta=\pi$
but with varying impurity scattering. The dependence of the zero-energy peak 
width with $\sqrt{\Gamma}$ is plotted
in the right inset of the left panel. In the panel to the right the impurity scattering is 
in the unitary limit. As expected the broadening of the Andreev states is very much suppressed.
}
\label{Spin_down_DOS}
\end{figure}
\end{minipage}
\begin{multicols}{2}
\be
N_{\uparrow} (\bvpf,\varepsilon;\Theta) = 
Im \Bigg\lbrack \frac{\varepsilon^R \cos(\Theta/2)+\Omega \sin(\Theta/2)}
                     {\Omega \cos(\Theta/2)  -\varepsilon^R \sin(\Theta/2)}\Bigg\rbrack.
\label{spindos}
\ee
For spin-down quasiparticles
$N_{\downarrow}$ reads the same after substitution $\Theta\rightarrow-\Theta$.
This density of states has Andreev bound states inside the gap, i.e. for $|\varepsilon|< \Delta$. These
states are located at $\varepsilon_{b,\uparrow(\downarrow)}=\pm \Delta \cos (\Theta/2)$, with
$+(-)$ for the spin-up (-down) branch. It is notable that the DOS (\ref{spindos}) has exactly the
same form as the DOS calculated for a S/F/S weak link, for which the angle
$\Theta$ is shown to depend on the thickness of and exchange field in the F-layer\cite{RAD99b}.

The existence of bound states will lead to a reduction
of the order parameter amplitude
in the vicinity of the magnetic interface. This is seen in Figure 2 in Ref. [\onlinecite{TSR88}].
The pairbreaking occurs gradually as the bound state on either spin branch 
is tuned towards $\varepsilon_b=0$ with $\Theta\rightarrow\pi$. At $\Theta=\pi$ 
the order parameter is totally reduced at the interface and recovers to
its bulk value over a distance of order $\xi_o$, the zero temperature
coherence length in the superconductor. It turns out that the position in energy
of the surface states, $\varepsilon_b$, is not to sensitive to a spatial variation
in the order parameter.
The Andreev bound states would be delta peaks if 
the superconductor had an infinite mean free path. The presence of impurities in the bulk
give rise to a finite life time $\tau$ which
broadens the Andreev peak. The broadening of the Andreev states is sensitive to the scattering strength
of the impurities. In the main panels of the figures \ref{Spin_down_DOS} 
the DOS is shown for different values of spin-mixing $\Theta$.
The surface order parameter is self consistently
determined for each spin-mixing angle. 
The scattering rate is $\Gamma=1/2\tau=0.01$ in units of
$2 \pi T_c$   
and corresponds to a mean free path $\ell_{\rm{mfp}}=50\,\xi_o$. From the literature
on d-wave superconductors it is known that surface bound states are broadened by impurity scattering
as $\sim\sqrt{\Delta \Gamma}$ in the Born limit \cite{poeni99}. In the two insets in figure \ref{Spin_down_DOS}
the dependence of the width of the zero energy peak with scattering rate for $\Theta=\pi$ is shown.
For small scattering rates indeed the $\sqrt{\Delta \Gamma}$-dependence is recovered. If, on the other
hand, the scattering strength of the impurities is in the strong scattering limit the broadening of the
Andreev bound states will be exponentially small, $\sim \sqrt{\Delta \Gamma}\exp[{-\Delta/\Gamma}]$
[\onlinecite{poeni99}].
The effect of unitary scatterers is shown in the figure to the right in Fig. \ref{Spin_down_DOS}.

\section{Josephson current-phase relation and the energy state of the junction.} 
\label{JCPR}
Next, let us consider the Josephson current-phase relation
through a magnetically active point contact.
Assuming a point contact allows several simplifications.
Effects of the contact itself on superconductivity i.e. the order parameter profile may be disregarded. 
This holds true if the 
contact radius is taken much smaller than the superconducting coherence length. 
Furthermore, spin-neutral surface scattering alone does not affect 
an isotropic s-wave superconductor and thus bulk values of the coherence functions $\gamma,\tamma$ can be
used for the in-scattering ones in the boundary condition (\ref{boundarycondition}). 
An additional advantage of the point contact condition is that the
results will not depend on the presence of non-magnetic bulk impurities as the current
through the point contact depends only on bulk coherence functions.
On the other hand, the point
contact itself is fully described by its transmission $t$ and degree
of spin mixing, $\Theta$.  
The Josephson current through the contact is calculated as a
function
of the phase difference, $\phi$, between two superconductors by 
the current formula
\be
\bvj(\phi)=e N_f \int^\infty_{-\infty} \frac{d\varepsilon}{8\pi i} 
{\rm Tr} \langle \bvvf \hat \tau_3 \hat g^{K}_\phi(\bvpf,0;\varepsilon)\rangle_\bvpf.
\label{current}
\ee
Here $\hat g^{K}_\phi(\bvpf,0;\varepsilon)=(\hat g^{R}_\phi(\bvpf,0;\varepsilon)-\hat g^{A}_\phi(\bvpf,0;\varepsilon))
\tanh(\varepsilon/2T)$ is the equilibrium Keldysh Green function constructed from
the retarded and advanced ones, $\hat g^{R,A}_\phi(\bvpf,0;\varepsilon)$, at the interface.
Functions $\hat g^{R,A}_\phi(\bvpf,0;\varepsilon)$ are calculated at the interface so
the boundary condition (\ref{boundarycondition}) is fulfilled. 
The resulting critical current of the junctions is characterized by different transparency and spin-mixing 
angle and show a rich variety as seen in Figures \ref{CriticalCurrents}. 
This quantity is defined as the maximum amplitude of the current
reached between a phase difference of $0$ and $\pi$. The sign of the critical current is either
positive giving a $"0"$-junction or negative signaling a $"\pi"$-junction.  
For arbitrary transmission and spin mixing the analytic expression for the current is not very tractable
and numerical analysis of the current-phase relation is more practical. In the two extreme limits
of tunneling, $t << 1$, and high transparency, $t\simeq 1$, the analytical expression is simpler and reveals the
physics going on. 

Starting with the tunneling limit, the transmitted coherence functions $\Gamma,\Tamma$ as given by equations
(\ref{boundarycondition}) are expanded to
leading order in transparency ${\cal T}=|t|^2$. The expanded $\Gamma,\Tamma$ are then put into the
expression for $\hat g^R$, eq. (\ref{Pspinpropagator}), and to first order in ${\cal T}$ the
resulting current is given as 
follows  
\be
j(\phi;\Theta)= {\cal T} j_o \sin \phi\, \int^\infty_{-\infty}\frac{d\varepsilon}{4\pi} 
 \bigg\lbrack {\cal K}_+(\varepsilon,\Theta)+{\cal K}_-(\varepsilon,\Theta)\bigg\rbrack
\label{tunnelcurrent}
\ee
where ${\cal K}_{\pm}(\varepsilon,\Theta)= 
\lbrack \Omega \cos (\Theta/2) \pm \varepsilon \sin(\Theta/2)\rbrack^{-2} \tanh (\frac{\varepsilon}{2T})$ and
$j_o=2 e v_f N_f \Delta^2$.
In general the current is totally governed by the bound states at $\varepsilon_b=\pm\Delta \cos(\Theta/2)$
and their population at the given temperature $T$. 
Notable is that for all values of $\Theta$, the current-phase 
relation is sinusoidal.
Setting $\Theta=0$ reproduces the usual Ambegaokar-Baratoff expression \cite{AB63}. 
At $\Theta=\pi$, $\varepsilon_b$ is a zero-energy bound state and give 
rise to a $"\pi"$-junction with  a critical current which 
increases as $T^{-1}$ with decreasing temperature as shown in the inset in the left figure 
\ref{CriticalCurrents}. A similar anomaly in the critical current 
occurs for d-wave superconductors
\cite{BBR96}. 
The difference between the anomaly in the two types
of superconductors is that for d-wave superconductors any concentration of bulk impurities will give 
a finite width of the zero energy bound states and reduce the $T^{-1}$-anomaly. For 
an s-wave superconductor
at a point contact only inelastic scattering processes, phase-breaking impurities or, as shown below, a
finite ${\cal T}$ can give a similar broadening and the anomaly 
is quite robust due to long inelastic (phase-breaking) scattering times, i.e. $1/2\tau_{inel,(phase)}<<\Delta$. 
Going away from $\Theta=\pi$, $\varepsilon_b$ moves to finite energy and  the functions
${\cal K}_{\pm}(\varepsilon,\Theta)$ acquire double pole structure at $\varepsilon_b$. The two poles
are slightly shifted and have slightly asymmetric magnitude in residues. Given that the
residues also are of different sign, they contribute oppositely to the critical current. 
The separation in energy of the poles is dependent
on the imaginary part, $\delta$, of the energy $\varepsilon+i\delta$. $\delta$ can loosely be 
interpreted as an inelastic scattering rate.
At $T=0$ and $\delta\rightarrow 0$ it turns
out that the
ground state of the junction is always a "0"-junction except at $\Theta=\pi$. As temperature and/or
the $\delta$ is increased the region in $\Theta$ showing a $"\pi"$-state junction
increases. This is clearly
demonstrated in figure \ref{CriticalCurrents} where the critical current is plotted as function of temperature.
For all but the two largest values of $\Theta$ the low-T critical current is positive.

Sticking to the tunneling limit, the results derived here can also be obtained using equations (93-94) of MRS.
What is crucial to note is that in order to find the bound state contribution to the current the Green's functions 
must be the ones arrived at in eq. (\ref{Pspinpropagator}). This is taking into account the spatial dependence
of the Green's function at the magnetic pinhole, i.e. solving the impenetrable wall problem with
the $\hat S$-matrix describing the pinhole. If this spatial dependence is neglected, and the bulk Green's
functions are used in equations (93-94) of MRS, the resulting
Josephson current will simply have two contributions $\sin(\phi\pm\Theta)$, one contribution for each spin band.

Moving away from the tunneling limit the $T^{-1}$-anomaly is cut off by the finite value of ${\cal T}$.
Instead, the switching between "0" and $"\pi"$-states happens in abrupt jumps in
the critical current.  This abruptness is spurious since the current-phase relation has three zeros
between 0 and $\pi$ and the change of state from $"\pi"$ to "0"
occurs without the current being zero for every phase difference. Instead the junction has two
local minima in energy, at phase difference $\phi=0$ and $\pi$. Right at the switching point the
two states are degenerate and the junction state can be tuned with temperature. This is shown in
the inset of the middle figure \ref{CriticalCurrents} which depicts the junction energy vs. phase
difference for a junction with ${\cal T}=0.1$ and $\Theta=3\pi/4$. 
As seen, between the two energy minima there is a
potential barrier. This barrier
is at highest at an intermediate phase $0 < \phi < \pi$ where the current through the junction is zero.
As temperature is swept over the switching temperature, which for this junction is at 
$T_{sw}\approx 0.12 T_c$,
the energy minimum jumps from $\phi=\pi$ to $\phi=0$ as temperature is 
increased through $T_{sw}$ and vice versa as 
the temperature is decreased through $T_{sw}$. 
The position in temperature depends on the two junction parameters
${\cal T}$ and $\Theta$. For larger transparencies the values of tunable junctions is restricted 
to a decreasing range in $\Theta$ just below $\Theta=\pi/2$. 

If the transparency is taken to unity 
the tunability of the junction state with temperature vanishes. 
This is seen from the current-phase relation
\be
j(\phi;\Theta)= j_o \int^\infty_{-\infty}\frac{d\varepsilon}{4\pi}
\bigg\lbrack {\cal J}_+(\phi;\varepsilon,\Theta)+{\cal J}_-(\phi;\varepsilon,\Theta)\bigg\rbrack
\label{pinholecurrent}
\ee
with 
\[
{\cal J}_\pm(\phi;\varepsilon,\Theta)={{\sin(\phi\pm\Theta)}\over{
\lbrack\Omega^2-\varepsilon^2+\Delta^2\cos(\phi\pm\Theta)\rbrack}}
\tanh (\frac{\varepsilon}{2T}).\]
The current is now controlled by interface states located at 
$\varepsilon_i=\pm \Delta\cos[(\phi\pm\Theta)/2]$, i.e. at a position given by the phase 
$\phi$ but shifted by $\pm\Theta$ for the two spin bands as compared to the spin-neutral case. 
This shows up in the fact that at  $\Theta=0$ the usual Kulik-Omel'yanchuk (KO) 
formula is recovered \cite{KO77} and 
at finite $\Theta$, eq. (\ref{pinholecurrent}) is a sum
of two KO supercurrents evaluated at phases shifted by $\pm\Theta$. 
For spin-mixing angles 
$\Theta< \pi/2$ the junction is in the "0"-state
and at $\Theta > \pi/2$ in the $"\pi"$-state. At $\Theta=\pi/2$ the junction state is degenerate for every
temperature as the current-phase relation has doubled periodicity.
\end{multicols}
\begin{minipage}[t]{0.95\textwidth}
\begin{figure}
\centerline{
\begin{minipage}[t]{.38\textwidth}
\centerline{\epsfxsize=0.8\textwidth{\epsfbox{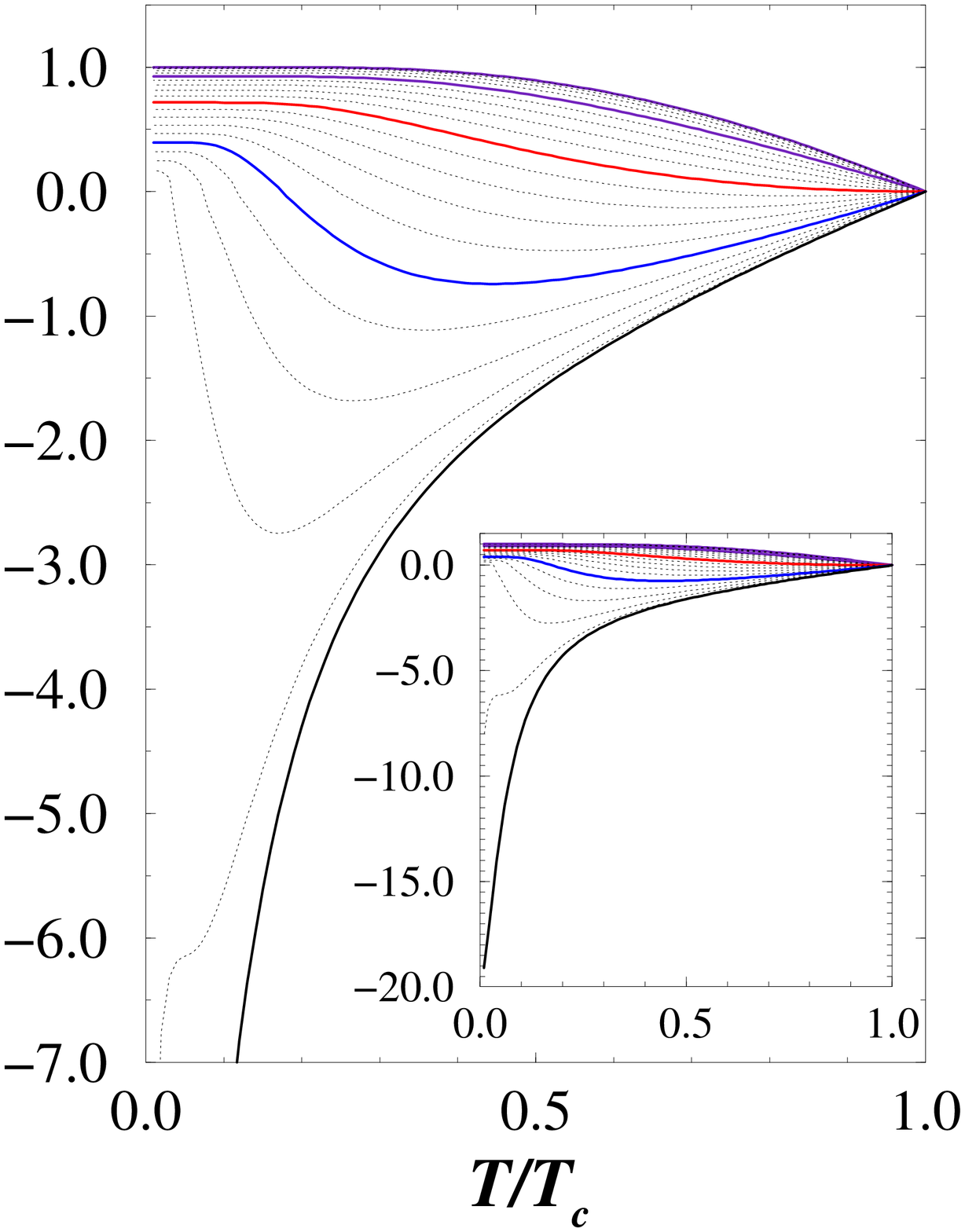}}}
\end{minipage}\hspace*{-.4truecm}
\begin{minipage}[t]{.38\textwidth}
\centerline{\epsfxsize=0.8\textwidth{\epsfbox{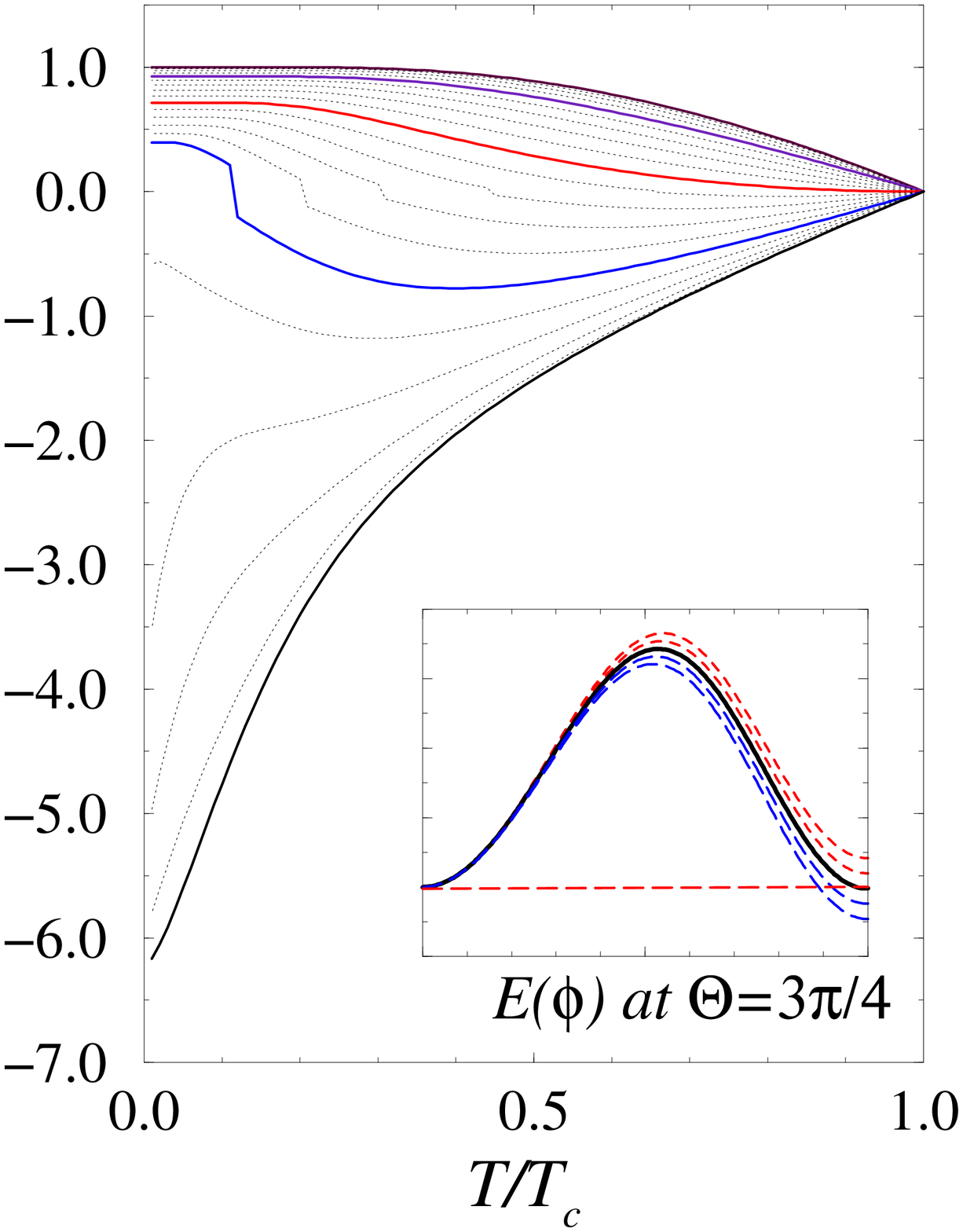}}}
\end{minipage}\hspace*{-.4truecm}
\begin{minipage}[t]{.38\textwidth}
\centerline{\epsfxsize=0.8\textwidth{\epsfbox{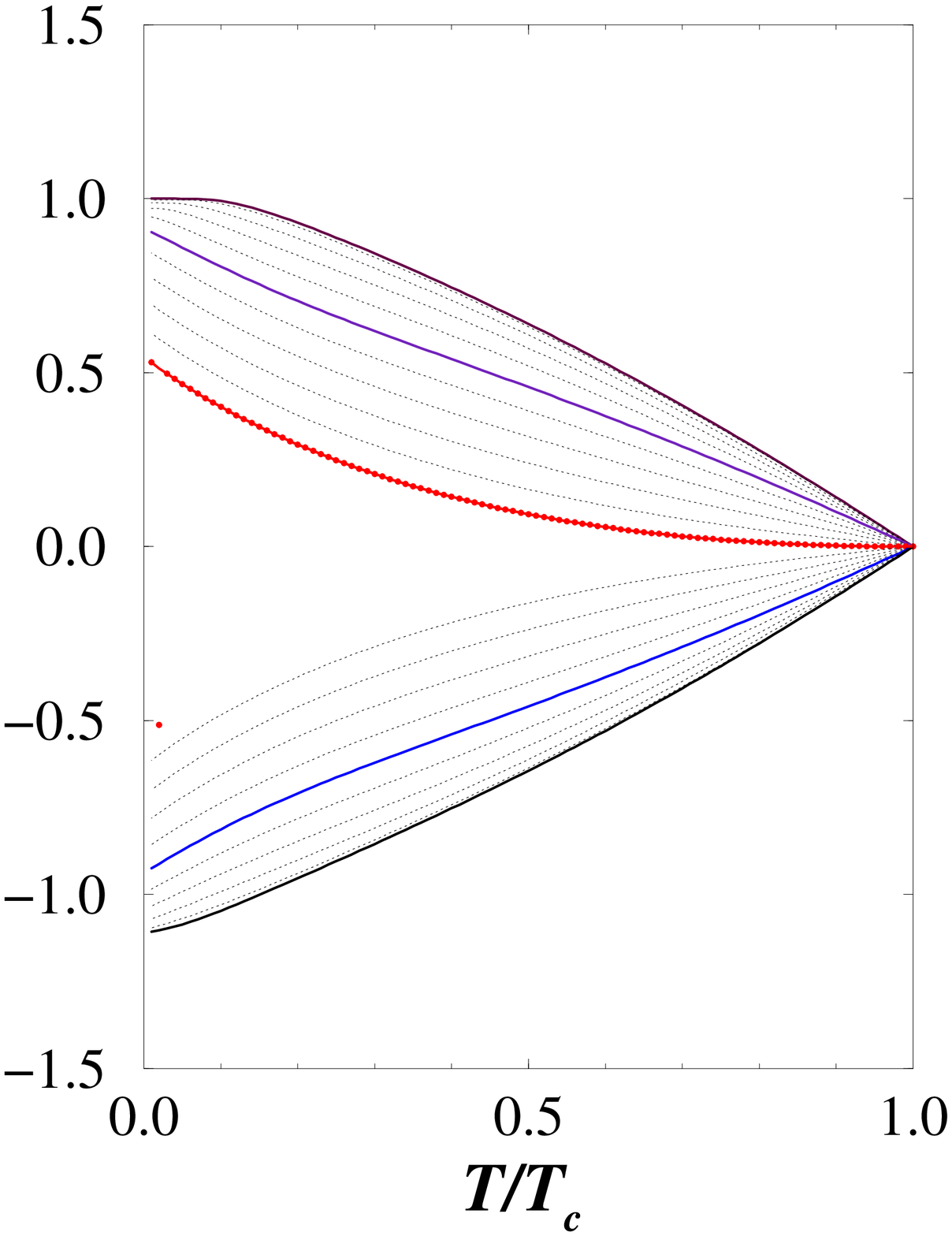}}}
\end{minipage}
}
\vspace*{0.5truecm}
\caption[]{
Critical currents for different transparencies $0.01, 0.1$ and $0.99$, left to right,
and for a dense sampling
of the spin-mixing angle, $\Theta$ running from 0 to $\pi$ in steps of $\pi/20$.
Thick lines are in intervals of $\pi/4$ as a guide for the eye.
All currents are scaled with the value of the critical current at $T=0$ and $\Theta=0$
for the current value of ${\cal T}$.
As is seen for all values of transparency, ${\cal T}$, the junction may either be a 
"0" or a "$\pi$"-junction
depending on the degree of spin mixing. In the low transparency limit and at $\Theta=\pi$ the
zero-energy bound state gives rise to a critical current $\sim T^{-1}$ as seen in the inset
of the left most panel. At intermediate $\Theta$ the
junction may switch between the "0" and the "$\pi$"-junction state with temperature. 
At larger ${\cal T}$ this switching becomes 
more abrupt in temperature. Finally, in the high transparency limit the switching 
between "0" and the "$\pi$"-junction state is lost and $\Theta$ defines the junction state 
for all temperatures.}
\label{CriticalCurrents}
\end{figure}
\end{minipage}
\begin{multicols}{2}

\section{Discussion}
In this paper a general solution is derived for the equilibrium part of the Zaitsev-Millis-Rainer-Sauls
boundary condition describing spin-active interfaces. This solution is the main result of the paper and
will be an important part in further studies of hybrid superconductor-(ferro)magnetic systems. As an
application, the effects of a magnetically active barrier, as described by a simple two-parameter $\hat S$-matrix,
are studied. In particular, it is shown that spin-mixing brings about Andreev bound states within the 
superconducting gap $\Delta$. The energy of these states is sensitive to the amount of
spin-mixing imposed by the scattering off the interface. 
Comparing with the DOS calculated here and those obtained in the tunneling experiments
of Stageberg {\it et al}\cite{STA85} it is plausible to conclude that the spin-mixing angle 
is not that large but rather
in the range
$|\Theta|\lesssim \pi/4$ for the materials studied in the experiment\cite{STA85}. 
None the less, it is important to note that the shift seen in the tunneling
conductance may be dependent on, and described by, 
the tunneling barrier properties\cite{TSR88} and thus not directly dependent of the
exchange field in the ferromagnet probed\cite{MT94,alex85}. On account of the Josephson coupling through a 
magnetically active interface, a small value of $\Theta$ would imply that "$\pi$"-junctions are hard to realize
at least in large junctions. To obtain
a "$\pi$"-junction it is shown that $\Theta$ must exceed $\pi/2$ for any range of transparency. On the other 
hand,
new experiments are in the making like the magnetic Cobalt grains studied by S.~Gu\'{e}ron 
{\it et~al}\cite{Gueron99}.
These small magnetic systems may well prove to offer magnetic scattering
where a large $\Theta \sim\pi$ is realized. As an example, using STM-techniques, 
as those performed on the Au point
contacts in Ref. [\onlinecite{Scheer98}], on small 
magnetic grains and with superconducting electrodes,
the Josephson physics described in this paper could be probed.

\acknowledgements
It is a pleasure to thank Yu.~S.~Barash, J.~C.~Cuevas, M.~Eschrig, T.~T.~Heikkil\"a, 
G.~Sch\"on and A.~Shelankov for stimulating discussions and comments
regarding topics of this work. Computer resources at Center for Scientific Computing
in Esbo, Finland are greatfully acknowledged. 
This work was supported by DFG project SFB 195. 

\appendix
\section{Quasiclassical Theory}
\label{QCT}
Calculations presented in this paper are done within the quasiclassical approximation
which is a generalization of the Landau Fermi-liquid theory 
to include superconduncting\cite{eilenberger68}
and superfluid\cite{SR83} phenomena. Quasiclassical theory 
is an expansion in quantities like
$T/T_f$ or $1/ \xi k_f$, which are usually of order $\sim 10^{-2}-10^{-3}$ 
in conventional superconductors.
I use the  quasiclassical theory for a p-wave superfluid $^3$He as worked
out by Serene and Rainer\cite{SR83} together with the
real-metal-oriented weak-coupling theory of Alexander {\it et~al}\cite{alex85}
to describe the superconductor in proximity to a magnetically active material.
In this appendix a  brief review, or collection, of the building blocks of 
quasiclassical theory is given.

Our starting point is the Eilenberger equation
\be
\qcgrad \hat g
+\lbrack \varepsilon\hat \tau_3-\hat v
-\hat \Sigma,
\hat g\rbrack =0
\label{Eilenberger}
\ee
for the $4\times 4$ matrix propagator $\hat g(\bvpf,\bvR;\varepsilon)$.
Here $\bvvf$ is the Fermi velocity, $\bvpf$ is a point on the Fermi surface,
The explicit $2\times2$-matrix structure of $\hat g$ reflects particle-hole (Nambu) space.
The spin degree of freedom is in the
parameterization into spin scalars, $g(\bvpf,\bvR;\varepsilon)$, 
$\tilde g(\bvpf,\bvR;\varepsilon)$, $f(\bvpf,\bvR;\varepsilon)$, 
$\tilde f(\bvpf,\bvR;\varepsilon)$, and spin
vectors, $\gv(\bvpf,\bvR;\varepsilon)$, $\tilde \gv(\bvpf,\bvR;\varepsilon)$, 
$\fv(\bvpf,\bvR;\varepsilon)$, $\tilde \fv(\bvpf,\bvR;\varepsilon)$ as
\be
\hat g=
\left(
\begin{array}{cc}
g+\gv\cd\bsp&
(f+\fv\cd\bsp)i\sigma_2 \\ 
i\sigma_2(\tilde f-\tilde \fv\cd\bsp)&
\sigma_2(\tilde g-\tilde \gv\cd\bsp)\sigma_2
\end{array}\right).
\label{matrixg}
\ee
In addition to (\ref{Eilenberger}) 
the propagator obeys the normalization condition $\hat g^2(\bvpf,\bvR;\varepsilon)=-\pi^2$.
There is some redundancy in the parameterization of (\ref{matrixg}) 
which gives the following symmetries\cite{SR83}
\be
\tilde x(\bvpf,\bvR;\varepsilon)= x(-\bvpf,\bvR;-\varepsilon^{*})^{*}
\label{symmetries}
\ee
where $x$ ($\tilde x$) is one of the spin components $g_{\alpha\beta}$ or $f_{\alpha\beta}$
 ($\tilde g_{\alpha\beta}$ or $\tilde f_{\alpha\beta}$) of the Green's function.
Matsubara propagators are obtained by $(\varepsilon\rightarrow i\varepsilon_n=i\pi T(2n+1))$,
retarded propagators
by $(\varepsilon\rightarrow \varepsilon+i\delta)$, and advanced propagators 
by $(\varepsilon\rightarrow \varepsilon-i\delta)$.
Analogous symmetry relations hold for the self energies.

The self energy $\hat \Sigma$ in equation (\ref{Eilenberger}) contains 
impurity contributions, the Fermi-liquid mean fields and the order parameter
\be
\hat \Sigma(\bvpf,\bvR;\varepsilon)=
\hat \Sigma_{\rm{i}}(\bvpf,\bvR;\varepsilon)
+\hat \Sigma_{\rm{m}}(\bvpf,\bvR) +\hat \Delta(\bvpf,\bvR)
\ee
The self-consistency equations for the
impurity self energy $\hat \Sigma_{\rm{i}}$ given one impurity potential $\hat u_{\rm{i}}$ and 
one impurity
concentration $n_{\rm{i}}$ is
\be
\hat \Sigma_{\rm{i}}(\bvpf,\bvR;\varepsilon)=n_{\rm{i}} \hat t_{\rm{i}}(\bvpf,\bvpf,\bvR;\varepsilon) \\ 
\label{impurities}
\ee
with the quasiclassical T-matrix equation
\be
\begin{array}{l}
\hat t_{\rm{i}}(\bvpf,\bvpf^\prime,\bvR;\varepsilon)=\hat u_{\rm{i}}(\bvpf,\bvpf^\prime)\\*[0.2truecm]
\quad+N_f \langle \hat u_{\rm{i}}(\bvpf,\bvpf^{\prime\prime}) 
\hat g(\bvpf^{\prime\prime},\bvR;\varepsilon) 
\hat t_{\rm{i}}(\bvpf^{\prime\prime},\bvpf^\prime,\bvR;\varepsilon)\rangle_{\bvpf^{\prime\prime}}
\end{array}
\label{tmatrix}
\ee
$N_f$ is the averaged normal state
density of states at the Fermi surface and
$\langle \dots \rangle_{\bvpf}$ denotes a Fermi-surface average. If there are more than one type
of impurity potentials equation (\ref{impurities}) will be a sum over the different impurity
contributions, each with its own density and its own T-matrix.
It is important to bare in mind that $\hat \Sigma_{\rm{i}}$ is in general not diagonal in 
particle-hole space.
The self energy $\hat \Sigma_{\rm{m}}$ contains the Fermi-liquid mean-field self energies.
It is diagonal in particle-hole space and divided into 
a symmetric ($\Ss$) and an antisymmetric ($\St$) part as
\ber
\Ss(\bvpf,\bvR)=T\sum_{\varepsilon_n} 
\big\langle  A^s(\bvpf,\bvpf^\prime) g(\bvpf^\prime,\bvR;\varepsilon_n) \big\rangle_{\bvpf^\prime}\label{landau}\\
\St(\bvpf,\bvR)=T\sum_{\varepsilon_n} 
\big\langle A^a(\bvpf,\bvpf^\prime) \gv(\bvpf^\prime,\bvR;\varepsilon_n) \big\rangle_{\bvpf^\prime}\nonumber
\eer
The Fermi-liquid interactions
$A^{(s,a)}(\bvpf,\bvpf^\prime)$ 
are parameterized by the Fermi-liquid parameters $A^s$ 
and $A^a$ which are phenomenological parameters determined from experiments.
The order parameter $\hat \Delta$ is split
into a singlet $\Delta$ and a triplet $\Dt$ parts 
by the singlet and triplet pairing interactions $V^s(\bvpf,\bvpf^\prime)$
and $V^t(\bvpf,\bvpf^\prime)$, and is calculated as
\ber
\Ds(\bvpf,\bvR)=T\sum_{\varepsilon_n}\big\langle V^s(\bvpf,\bvpf^\prime)f(\bvpf^\prime,\bvR;\varepsilon_n)
\big\rangle_{\bvpf^\prime}\\
\label{singletop}
\Dt(\bvpf,\bvR)=T\sum_{\varepsilon_n}\big\langle V^t(\bvpf,\bvpf^\prime)\fv(\bvpf^\prime,\bvR;\varepsilon_n)
\big\rangle_{\bvpf^\prime}.
\nonumber
\eer

The set of equations written above, the Eilenberger equation for $\hat g$ and the  
equations for the self energies $\hat \Sigma$,
must be solved self-consistently by iteration together 
with the appropriate boundary conditions imposed on the propagator.
With $\hat g(\bvpf,\bvR;\varepsilon)$ determined, physical quantities like the current densitymay be computed 
\be
\jv(\bvR)=2e N_f T\sum_{\varepsilon_n} \big\langle \bvvf(\bvpf) g(\bvpf,\bvR;\varepsilon_n) \big\rangle_{\bvpf}.
\label{Matcurrent}
\ee
The local density of states resolved for a given $\bvpf$ and a given spin direction $\ev$ 
is calculated at real energies as
\be
N_{\ev}(\bvpf,\bvR;\varepsilon^R)
=-\frac{N_f }{\pi} {\rm Im}\bigg\lbrack g(\bvpf,\bvR;\varepsilon^R)+\ev\cd
\gv(\bvpf,\bvR;\varepsilon^R)\bigg\rbrack
\label{Espindos}
\ee

\end{multicols}


\begin{thebibliography}{10}
\bibitem{ph} 
Stability properties of the Riccati equations are closely related to
particle-like
and hole-like solutions of the Andreev equations \cite{OS00}. $\gamma$ can be seen as a
coherence function relating the Andreev amplitudes $(u,v)$ as $v=\gamma u$ were $(u,v)$
correspond to a particle like solution, i.e. with a group velocity along the 
$\bvp$. The same goes for $\tamma$ but 
$(u,v)$ corresponding to a hole-like solution as $u=\tamma v$.
\end{thebibliography}
\end{document}